\makeatletter
\def\ps@ieeesubmission{%
  \let\@mkboth\@gobbletwo
  \def\@oddhead{}%
  \def\@evenhead{}%
  \def\@oddfoot{%
    \parbox[b]{\textwidth}{%
      \normalfont\fontsize{7}{8}\selectfont
      This work has been submitted to the IEEE for possible publication.
      Copyright may be transferred without notice, after which this version
      may no longer be accessible.\par
    }%
  }%
  \let\@evenfoot\@oddfoot
}
\makeatother
\AddToHook{begindocument/end}{\thispagestyle{ieeesubmission}}
\documentclass{article}
\usepackage{spconf,amsmath,graphicx,hyperref}
\usepackage{booktabs}
\usepackage{tabularx}

\title{PI--AMFM: Permutation-Invariant Learning for Variable-Cardinality AM--FM Mode Decomposition in Biomedical Signal Analysis}
\name{Youngsun Kong and Ki H. Chon}
\address{Biomedical Engineering, University of Connecticut, Storrs, CT 06029, USA}
\begin{document}
\ninept
\maketitle
\begin{abstract}

Physiological recordings often contain nonstationary oscillatory components whose number and dynamics vary across signals.
Amplitude- and frequency-modulated (AM--FM) representations are well suited to characterizing such dynamics and have shown broad utility in biomedical signal analysis.
Recent approaches have incorporated neural networks to learn mode decomposition patterns from data, but component cardinality is often predefined or determined through separate stopping or selection mechanisms.
We propose a permutation-invariant neural framework for variable-cardinality AM--FM mode decomposition (PI--AMFM).
PI--AMFM combines a multiscale temporal encoder, Mamba backbone, and component-presence estimation, with permutation-invariant Hungarian matching during training.
On synthetic AM--FM signals, PI--AMFM achieved lower decomposition, instantaneous-frequency, reconstruction, and mode-count errors than the compared methods
while preserving the overall trajectory pattern in a crossing-chirp example.
On photoplethysmographic recordings, recovered modes captured cardiac and respiratory dynamics despite training only on synthetic signals.
These results support the feasibility of PI--AMFM for variable-cardinality decomposition of nonstationary biomedical signals.

\end{abstract}
\begin{keywords}
AM--FM mode decomposition, nonstationary signal processing, permutation-invariant learning, Mamba, photoplethysmography
\end{keywords}
\section{Introduction}
\label{sec:intro}

Many physiological recordings can be viewed as mixtures of nonstationary oscillatory components arising from multiple physiological processes operating at different and time-varying temporal scales.
Photoplethysmography (PPG) provides a representative example, with a dominant cardiac oscillation accompanied
by respiratory modulation through changes in amplitude, baseline, and pulse timing.
Physiological variability, sensor noise, and motion artifacts further complicate the separation of these underlying components.
These characteristics motivate decomposition methods capable of representing oscillations with time-varying amplitudes and frequencies.

An amplitude- and frequency-modulated (AM--FM) representation provides a natural description of such components through their time-varying amplitude and instantaneous frequency (IF).
Several AM--FM-based approaches have been developed for nonstationary signal analysis and decomposition.
Variable-frequency complex demodulation (VFCDM) provides high-resolution characterization of time-varying amplitude and IF \cite{wang2006high},
and has demonstrated strong performance in biosignal applications \cite{posada2016highly,kong2025new}.
Variational nonlinear chirp mode decomposition (VNCMD) models wideband AM--FM modes with adaptively estimated IF trajectories \cite{chen2017nonlinear}
and has demonstrated strong performance in EEG analysis \cite{kamble2023novel};
however, the number of modes must be specified in advance.
Adaptive chirp mode pursuit (ACMP) further enables recursive extraction of AM--FM modes without prior knowledge of their number \cite{chen2019adaptive}.
More recently, PDS--SSA \cite{zhang2025parameterized} combines parameterized demodulation with sliding singular spectrum analysis to decompose multicomponent AM--FM signals without requiring either the number of components or prior IF trajectories.
ADIN \cite{zhang2026adaptive} embeds demodulation, adaptive filtering, and
re-modulation within a neural architecture for multicomponent AM--FM separation,
but assumes a predefined component count and prior estimates of the component IF trajectories.

Among existing AM--FM decomposition approaches,
adaptive model order is commonly achieved through iterative, signal-dependent extraction,
whereas neural-network (NN)-based formulations often assume a predefined component structure or prior frequency information.
We propose PI--AMFM, a permutation-invariant NN framework for decomposing a signal into an unknown number of time-varying AM--FM modes
without imposing predefined IF trajectories or extraction order.
PI--AMFM aims to combine the representational capacity of learning-based methods
with direct, parallel prediction of a variable number of AM--FM modes using a pretrained permutation-invariant model.

\section{Methodology}
\label{sec:format}

\subsection{PI--AMFM Architecture}
\label{ssec:architecture}

The PI--AMFM architecture comprises
a multi-scale temporal encoder, a Mamba backbone \cite{gu2023mamba}, and a slot-based output layer (Fig.~1).
The multi-scale encoder uses parallel convolutions with kernel sizes $k\in\{3,7,15\}$ to capture temporal patterns at different scales,
followed by six bidirectional Mamba blocks for temporal modeling.

The output layer contains $K=8$ candidate decomposition slots, selected empirically.
Each slot produces two time-series outputs corresponding to the real
and imaginary components of a predicted AM--FM mode, together
with a sigmoid-transformed presence score $p_k \in [0,1]$.
Slots with $p_k \geq 0.5$ were classified as active and included in reconstruction;
the estimated mode count was $\hat{M}=\sum_{k=1}^{K}\mathbf{1}[p_k\geq0.5]$.
The framework supports up to eight active modes without requiring the component count at inference.

Because the ordering of the underlying components is arbitrary,
no fixed correspondence is imposed between a physiological or frequency component and a particular output slot.
Inspired by prior permutation-invariant set-prediction approaches \cite{houba2025slotflow},
Hungarian matching \cite{kuhn1955hungarian} was used during training
to determine the minimum-cost assignment between predicted slots and ground-truth modes.
Unmatched slots are treated as inactive components.

\begin{figure}[t]
    \centering
    \includegraphics[width=\columnwidth]{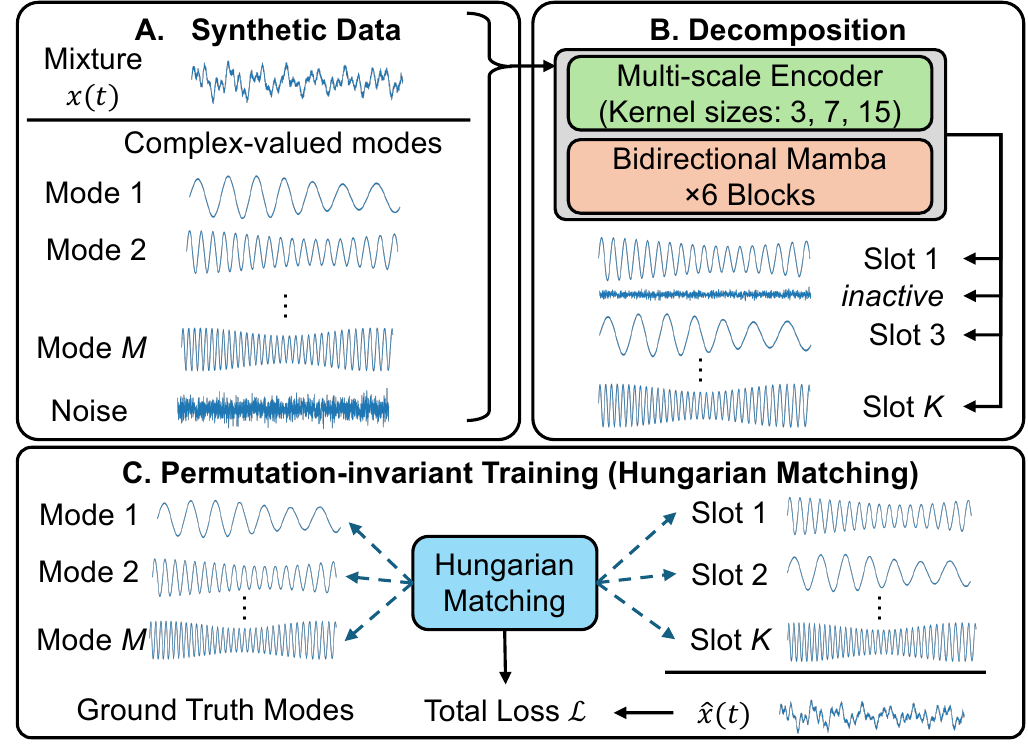}
    \caption{Overall PI--AMFM architecture and permutation-invariant training strategy with $K=8$ candidate slots.}
    \label{fig:results}
\end{figure}

\subsection{Synthetic AM--FM Data and Model Training}
\label{ssec:training}

PI--AMFM was trained on synthetic multicomponent AM--FM signals with known component-level ground truth,
with the number of active components $M$ randomly selected from 1 to 8.
Each mode and the real-valued mixtures were represented as
\begin{equation}
\begin{aligned}
z_m(t) &= A_m(t)e^{j\phi_m(t)}, \qquad
f_m(t)=\frac{1}{2\pi}\frac{d\phi_m(t)}{dt}, \\
x(t) &= \sum_{m=1}^{M}\Re\{z_m(t)\}+\epsilon(t).
\end{aligned}
\label{eq:mode_and_mixture}
\end{equation}
where $\epsilon(t)$ denotes additive Gaussian noise scaled to a signal-to-noise ratio (SNR) sampled uniformly from $-5$ to $30$~dB.

Amplitude envelopes combined randomly scaled smooth cubic-spline modulation,
optional sinusoidal AM (50\% probability per mode), and optional temporal gating (30\% probability per mode).
IF trajectories were drawn with equal probability from the constant,
linear-chirp, quadratic-chirp, sinusoidal-FM, and smooth random-FM families.
Trajectory amplitudes were randomized within the allowable frequency range,
and phases were obtained by integrating the resulting IFs with random initial phase.
The complex-valued modes were retained as the decomposition targets.

Signal lengths were sampled between 512 and 4096 samples to accommodate varying record durations and frequency scales.
With normalized sampling frequency $f_s=1$, IFs were bounded between
0.005 and 0.45 cycles/sample.
Nominal component frequencies were constrained to contain at least four cycles
per record and to have a minimum pairwise separation of 0.005 cycles/sample;
instantaneous trajectories were nevertheless allowed to overlap or cross.
Signals were normalized to unit clean-mixture rms before noise addition.
We independently generated 100,000 training examples and 10,000 examples each
for validation and testing.

Inspired by variable-cardinality set prediction frameworks \cite{carion2020end}, the model was trained using the following composite loss function:
\begin{equation}
\mathcal{L}
=
\mathcal{L}_{\mathrm{mode}}
+ 0.15 \mathcal{L}_{\mathrm{presence}}
+ 0.05 \mathcal{L}_{\mathrm{recon}}
+ 0.05 \mathcal{L}_{\mathrm{inactive}}
\label{eq:total_loss}
\end{equation}
where $\mathcal{L}_{\mathrm{mode}}$ is the target-energy-normalized complex squared error over Hungarian-matched predicted and ground-truth modes.
$\mathcal{L}_{\mathrm{presence}}$ applies binary cross-entropy to classify matched and unmatched slots as active and inactive, respectively.
$\mathcal{L}_{\mathrm{recon}}$ is the mean absolute error (MAE) between the clean target mixture and the sum of the real parts of matched predicted modes.
$\mathcal{L}_{\mathrm{inactive}}$
is the mean squared value of the real and imaginary outputs of unmatched slots.

The model was trained for 50 epochs using AdamW with a learning rate of $3\times10^{-4}$,
weight decay of $10^{-4}$, and a batch size of 64.
A cosine schedule with 5\% warm-up was used,
with model selection based on validation performance.
The Mamba backbone comprised six bidirectional Mamba-1 blocks with a model dimension of 128, a state dimension of 16, and a dropout rate of 0.1.

\subsection{PPG Respiratory Data and Evaluation}
\label{ssec:ppg_evaluation}

For physiological evaluation, the BIDMC dataset was selected because PPG's cardiac and respiratory components can both exhibit time-varying amplitude and frequency \cite{pollard_physionet_2026,pimentel2016toward},
making it an optimal case. The dataset contains 53 eight-minute recordings from critically ill patients, with simultaneously acquired PPG and impedance-respiratory signals sampled at 125~Hz.
Reference heart rate (HR) derived from the electrocardiogram (ECG) is also provided. Respiratory rate (RR) was derived from breaths manually annotated by two individuals.
For the present analysis, the PPG signals were resampled to 10~Hz using an anti-aliasing filter to reduce computational complexity while preserving the cardiac and respiratory frequency ranges of interest.
The resampled signals were median-centered to remove the baseline offset and normalized by their rms before decomposition. No bandpass filtering was applied.

\subsection{Benchmarks}
\label{ssec:benchmarks}

Because PI--AMFM targets decomposition without prior knowledge of component cardinality or frequency trajectories, methods that operate without such information were compared.
Among AM--FM approaches, DD--ACMD was selected because it satisfies these criteria and has a reproducible implementation \cite{wang2023data,wang2026pysdkit}.
We also compared IDSD \cite{huijben2026idsd}, an NN-based method that performs sequential frequency-domain extraction and determines cardinality using a residual-based stopping criterion.
The authors' released implementation was adapted to our variable-length synthetic data and augmented with the stopping criterion described in the paper \cite{huijben2026idsd}.
IDSD was retrained on the same synthetic dataset using the PI--AMFM training optimization protocol while retaining its original loss.
IDSD retrained using its original protocol performed worse on all metrics except retained-count MAE (1.788 vs.\ 1.809).
We also included EMD \cite{huang1998empirical}, CEEMDAN \cite{torres2011complete} using 100 ensemble trials,
and FIF \cite{cicone2021numerical} as classical decomposition benchmarks. All comparison methods used an extraction cap of 20 modes to balance decomposition capacity and computational cost.

\subsubsection{Synthetic Signal Decomposition}
\label{sssec:benchmark_synthetic}

PI--AMFM directly produced complex-valued modes; the other benchmark modes were converted to analytic signals using the Hilbert transform.
For the comparison methods, the $K$ highest-energy modes were retained for quantitative evaluation when more than $K$ modes were extracted.
Hungarian matching used normalized complex squared error between the $\hat{M}$ retained modes and the $M$ true modes.
If $\hat{M}<M$, zero-valued placeholders were appended before matching; if $\hat{M}>M$, unmatched modes were excluded from matched waveform and IF errors but retained for mode-count and reconstruction metrics.

For each matched pair, normalized complex MSE (NMSE) was calculated by normalizing the squared error by the reference-mode energy, and the per-signal mode NRMSE was defined as
\begin{equation}
\mathrm{NRMSE}_{\mathrm{mode},s}
=
\sqrt{\frac{1}{M_s}\sum_{m=1}^{M_s}\mathrm{NMSE}_{s,m}},
\label{eq:NRMSE}
\end{equation}
where $M_s$ is the number of true modes in signal $s$.

IF was computed from adjacent-sample phase increments:
\begin{equation}
\hat f[t]
=
\frac{1}{2\pi}
\arg\!\left(\hat z[t+1]\hat z^{*}[t]\right),
\label{eq:IFcomputation}
\end{equation}
and compared with the mean reference IF over the corresponding sample interval.
Unlike waveform error, which was evaluated over the full record,
IF error was evaluated only where both adjacent reference samples were active and exceeded $10^{-3}$ of the reference-mode peak amplitude, because IF is ill-defined near zero amplitude.
Missing modes were assigned zero IF.

Reconstruction NRMSE was computed against the clean mixture using all retained predicted modes, excluding residual or trend components.
All metrics were computed per signal and then averaged equally across the 10,000 test signals.

Crossing behavior was additionally evaluated using two equal-amplitude linear chirps of 2,048 samples with an initial phase difference of $\pi/2$.
The resulting noise-free mixture was normalized to unit rms.
Their IFs varied oppositely between 0.08 and 0.22 cycles/sample, crossing at 0.15 cycles/sample at the midpoint. For each method, the two highest-energy modes were displayed.

\subsubsection{PPG Respiratory Component Extraction}
\label{sssec:benchmark_ppg}

Cardiac- and respiratory-related modes were independently selected by maximum integrated Welch power within the cardiac (0.7--3.5~Hz) and respiratory (0.1--0.7~Hz) bands, respectively.
Although respiratory frequencies can approach 1~Hz under high ventilatory demand \cite{nicolo2020importance},
a 0.7-Hz upper bound was used to define adjacent selection bands for this feasibility analysis.
No constraint required different modes for cardiac and respiratory attribution, and reference measurements were not used for mode selection.
The analysis was intended to demonstrate physiologically meaningful mode recovery rather than general-purpose mode attribution.

Because PPG signals can contain prominent cardiac harmonics \cite{masinelli2021spare},
a simple harmonic rule was applied, with empirically set thresholds.
Among the eligible modes, the two modes with the highest cardiac-band power were examined when the second-ranked mode had at least 85\% of the power of the highest-ranked mode.
The candidate with the lower median interior IF was selected when the ratio of the higher to lower IF remained between 1.9 and 2.1 for at least 90\% of interior samples.
If these conditions were not satisfied, the highest-power mode was retained.

IFs were obtained from the unwrapped phase of each complex-valued PI--AMFM mode; for the other methods, analytic signals were constructed using the Hilbert transform.
Local phase slopes were estimated using centered least-squares fits over 8-s windows for RR and 4-s windows for HR. RR and HR estimates were clipped to 6--60 breaths/min and 42--210 beats/min, respectively, followed by a 3-s median filter.
The first and last 10~s were excluded, as they can contain incomplete RR cycles.
HR estimates were interpolated to the reference timestamps, whereas RR was averaged within each valid inter-breath interval and compared with $60/\Delta t_{\mathrm{breath}}$ for each annotator.
MAE and RMSE were calculated by pooling all eligible reference pairs.

\section{Results}
\label{sec:pagestyle}

\subsection{Synthetic Signal Decomposition}
\label{ssec:synthetic_results}

PI--AMFM correctly identified two active modes, whereas the compared methods
overdecomposed the signal into 4--20 extracted modes (Fig.~2).
Although PI--AMFM's IF estimates exhibited noticeable fluctuations, particularly early
in the signal and near the frequency crossing, PI--AMFM preserved the overall
two-trajectory crossing chirp pattern.
DD--ACMD followed the lower-frequency branch rather than a single chirp through the crossing.
IDSD followed locally ordered frequency branches rather than preserving
source identity through the crossing.
FIF failed to recover the lower trajectory, while EMD and CEEMDAN exhibited substantial
mode mixing and unstable IF estimates.

\begin{figure}[t]
\centering
\includegraphics[width=\columnwidth]{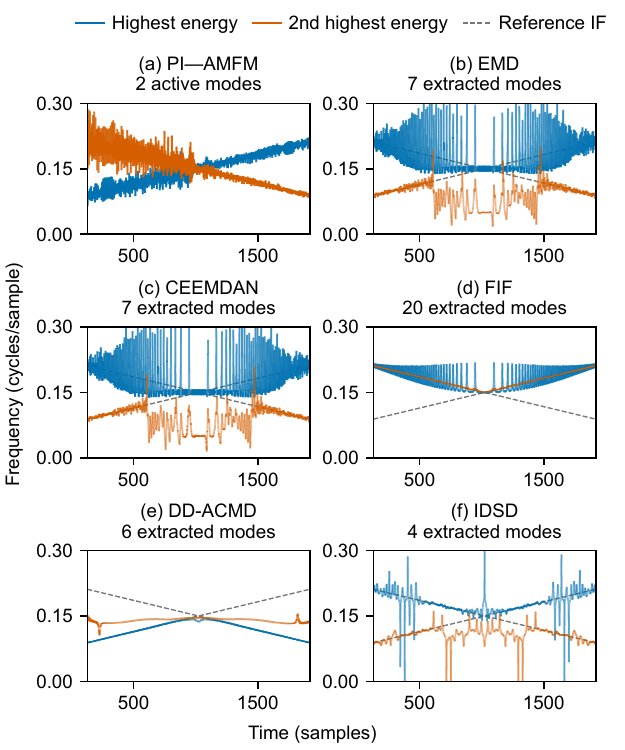}
\caption{Crossing-chirp decomposition. Higher opacity indicates greater relative mode amplitude within each panel.}
\label{fig:chirps}
\end{figure}

PI--AMFM's greater local variability in the estimated IF trajectories may reflect differences in inductive bias.
DD--ACMD, FIF, and IDSD use explicit frequency-selective decomposition mechanisms that may favor smoother modes and suppress local fluctuations.
In contrast, PI--AMFM allows greater flexibility in the recovered AM--FM representation, potentially improving source-specific trajectory recovery at the cost of weaker local IF smoothing.

PI--AMFM outperformed all compared methods, including IDSD retrained on the same training set, across all evaluated metrics on the in-distribution synthetic test set (Table~1).
Although this test follows the same stochastic signal model used for training, it provides controlled quantitative evaluation with component-level ground truth unavailable in real-world recordings.

\begin{table}[t]
\centering
\caption{Mean synthetic decomposition performance.}
\label{tab:synthetic_results}
\small
\setlength{\tabcolsep}{2pt}

\begin{tabularx}{\columnwidth}{@{}Xccccc@{}}
\toprule
\textbf{Method} & Complex & Real & IF & Recon. & Count \\
 & NRMSE & NRMSE & MAE & NRMSE & MAE \\
\midrule
PI--AMFM & \textbf{0.411} & \textbf{0.411} & \textbf{0.025} & \textbf{0.199} & \textbf{0.379} \\
EMD & 0.900 & 0.900 & 0.153 & 0.438 & 3.270 \\
CEEMDAN & 0.897 & 0.897 & 0.152 & 0.438 & 3.266 \\
FIF & 0.790 & 0.790 & 0.093 & 0.438 & 3.503 \\
DD--ACMD & 0.780 & 0.779 & 0.128 & 0.685 & 2.156 \\
IDSD & 0.621 & 0.620 & 0.067 & 0.495 & 1.809 \\
\bottomrule
\end{tabularx}

\vspace{2pt}
\raggedright
IF MAE: cycles/sample; Count MAE: retained-mode counts.
\end{table}

\subsection{PPG Respiratory Component Extraction}
\label{ssec:ppg_results}

PI--AMFM achieved the lowest RR error metrics (Table~2).
For HR, PI--AMFM was slightly outperformed by DD--ACMD, with small differences of 0.2~beats/min in MAE and 0.8~beats/min in RMSE.
DD--ACMD achieved the second-lowest RR errors, although its RMSE was approximately 2.7~breaths/min (33--34\%) higher than PI--AMFM's.
IDSD produced the highest RR errors, while achieving the third-lowest HR errors.
Fig.~3 illustrates PI--AMFM's recovered cardiac and respiratory IF trajectories.

\begin{table}[t]
\centering
\caption{Cardiorespiratory estimation on BIDMC.}
\label{tab:bidmc_results}
\small
\setlength{\tabcolsep}{2pt}

\begin{tabularx}{\columnwidth}{@{}Xcccccc@{}}
\toprule
\textbf{Method} & \multicolumn{2}{c}{RR Ann. 1} & \multicolumn{2}{c}{RR Ann. 2} & \multicolumn{2}{c}{HR} \\
\cmidrule(lr){2-3}\cmidrule(lr){4-5}\cmidrule(l){6-7}
(per min) & MAE & RMSE & MAE & RMSE & MAE & RMSE \\
\midrule
PI--AMFM & \textbf{4.8} & \textbf{8.1} & \textbf{4.9} & \textbf{8.1} & 1.6 & 4.1 \\
EMD & 9.1 & 13.4 & 9.0 & 13.2 & 8.4 & 19.0 \\
CEEMDAN & 7.7 & 11.0 & 7.7 & 11.0 & 7.9 & 18.2 \\
FIF & 9.2 & 12.5 & 9.1 & 12.4 & 4.9 & 16.7 \\
DD--ACMD & 5.6 & 10.8 & 5.6 & 10.8 & \textbf{1.4} & \textbf{3.3} \\
IDSD & 13.4 & 19.7 & 13.4 & 19.7 & 3.8 & 13.0 \\
\bottomrule
\end{tabularx}

\vspace{2pt}
\raggedright
\small
$n$ = 24,429 (HR), 6,934 (RR Ann.~1), and 7,014 (RR Ann.~2);
central 95\% reference RR ranges (breaths/min): 10.7--26.1 (Ann.~1), 11.1--26.8 (Ann.~2).
\end{table}

\begin{figure}[t]
\centering
\includegraphics[width=\columnwidth]{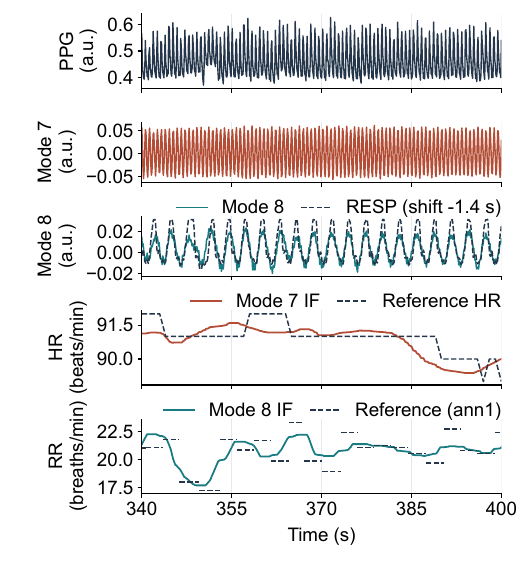}
\caption{PPG decomposition with PI--AMFM and IF-derived HR/RR measurements. RESP: impedance-respiratory signal.}
\label{fig:ppg}
\end{figure}

\section{Discussion and Conclusion}
\label{sec:typestyle}

Mamba was empirically selected as the temporal backbone after preliminary chirp-crossing comparisons with
residual convolutional neural networks (CNNs), long short-term memory networks (LSTMs), Transformers, temporal convolutional networks (TCNs), and the diagonal state-space model S4D.
One possible explanation is that Mamba's state-space formulation is well suited to modeling continuously evolving latent dynamics over time.
In particular, its input-dependent selective state-space mechanism enables the model to adaptively propagate
and update temporally relevant information across long sequences.

Many decomposition methods, including AM--FM-based and benchmarked approaches,
determine cardinality through iterative or successive component extraction and stopping criteria.
In our experiments, these methods exhibited overdecomposition and relatively high mode-count errors even after retaining at most $K$ modes,
and such stopping rules may require adjustment across signal conditions.
PI--AMFM instead estimates mode presence jointly across a fixed set of candidate slots,
avoiding sequential extraction and the associated propagation of errors through residual updates.
The tradeoff is that a maximum mode count must be predefined;
we used $K=8$ based on empirical observations.
A larger $K$ may increase ambiguity among candidate slots and make inactive-slot identification more difficult.

Many NN-based decomposition approaches assume a predefined number or structure of components, including
IRCNN \cite{zhou2024ircnn} and ADIN \cite{zhang2026adaptive}.
Other NN-based approaches explicitly address unknown cardinality.
IDSD addresses unknown cardinality through sequential extraction and residual-based stopping \cite{huijben2026idsd}.
SignalNet \cite{dreifuerst2022signalnet} predicts the sinusoid count using a classification module followed by a count-specific parameter estimator,
while SlotFlow \cite{houba2025slotflow} predicts a cardinality distribution and instantiates the corresponding number of permutation-invariant slots.
Both estimate fixed sinusoidal parameters rather than time-varying AM--FM waveforms.
PI--AMFM performs direct, parallel, permutation-invariant prediction of complex-valued time-varying modes with explicit per-slot presence estimation.

We selected PPG as a representative biosignal to demonstrate recovery of cardiorespiratory dynamics,
although HR and RR estimation from PPG are well-studied problems with many specialized methods.
The value of this example is therefore not to introduce a new method for cardiorespiratory rate estimation,
but to demonstrate that PI--AMFM can recover physiologically meaningful,
time-varying components from real signals despite being trained only on synthetic AM--FM mixtures.
The competitive performance of PI--AMFM and DD--ACMD, particularly for cardiac dynamics,
further supports AM--FM representations for physiological signal analysis.
Because many biosignals exhibit time-varying spectral dynamics, PI--AMFM may also be applicable to electrodermal activity \cite{posada2016highly},
ECG-derived skin nerve activity \cite{kong2025new}, and EEG \cite{kamble2023novel}.
Future work should evaluate these extensions and robustness across signal types and recording conditions.
Overall, the results support PI--AMFM as a learning-based framework for variable-cardinality decomposition of nonstationary physiological signals.

\section{Compliance with Ethical Standards}
This study used synthetic signals and public, de-identified BIDMC data from PhysioNet \cite{pollard_physionet_2026, pimentel2016toward}.
Use of this public dataset did not require institutional IRB review.

\section{Acknowledgements}
Youngsun Kong was supported by the National Institute of Dental and Craniofacial Research of the National Institutes of Health under Award Number F32DE033566. The authors declare no conflicts of interest.

\label{sec:refs}

\bibliographystyle{IEEEbib}
\bibliography{refs}

\end{document}